\documentclass[twocolumn,pra,showpacs,superscriptaddress]{revtex4}
\usepackage{amsmath}
\usepackage{amssymb}
\usepackage{amsbsy}
\usepackage{graphicx}
\usepackage[ansinew]{inputenc}

\begin{document}
\title{Thermalization and  Quantum Correlations in Exactly Solvable Models}
\author{Miguel A. Cazalilla}
\affiliation{Centro de F\'isica de Materiales CSIC-UPV/EHU. Paseo Manuel
de Lardizabal 5,  E-20018 San Sebastian, Spain}
\affiliation{Donostia International Physics Center (DIPC),
 Paseo Manuel de Lardizabal 4, E-20018 San
Sebastian, Spain}
\author{Anibal Iucci}
\affiliation{Instituto de F\'isica de La Plata (IFLP) - CONICET and Departamento de F\'isica,
Universidad Nacional de La Plata, cc 67, 1900 La Plata, Argentina}
\author{Ming-Chiang Chung}
\affiliation{Physics Division, National Center for Theoretical Science, Hsinchu 30013, Taiwan}
\affiliation{Institute of Physics, Academia Sinica, Taipei 11529, Taiwan}
\begin{abstract}
 The generalized Gibbs ensemble introduced for describing few body correlations in
exactly solvable systems following a quantum quench  is related to  the nonergodic way
in which operators sample, in the limit of infinite time after the quench, the quantum correlations present
in the initial state.  The nonergodicity of the correlations is thus shown \emph{analytically}
to imply the equivalence with  the generalized Gibbs ensemble  for quantum Ising and
 XX spin chains as well as for the Luttinger model the thermodynamic limit, and
for a broad class of initial states and correlation
functions of both  local and nonlocal operators.
\end{abstract}
\date{\today} \maketitle

\section{Introduction}

Following a number of groundbreaking experiments with ultracold atomic systems~\cite{Weiss,Schmiedmayer,
Bloch},  the problem of thermalization of exactly solvable quantum many-body
systems has attracted a great deal of
attention~\cite{Rigol_gge,Cazalilla,attention}. This is because
it relates to fairly fundamental questions, such as the emergence of
thermodynamics in isolated systems  prepared in initial states
that are not eigenstates of the Hamiltonian (i.e. systems undergoing 
a so-called `quantum quench').
The latter subject has  deep ramifications,
both in condensed matter physics~\cite{CazalillaRigol,Polkovnikovetal}
and cosmology~\cite{PolkovnikovGritsev,Polkovnikovetal}.
Moreover,  this problem is also relevant to the ongoing efforts
 to build `quantum emulators', that is, tunable quantum systems
 capable of accurately simulating the mathematical
models of many body physics. In this regard, the problem of thermalization
impacts on questions such like how much memory will the emulator
retain of its initial conditions and whether the standard Gibbsian ensembles
can be used to predict the outcome of the simulation~\cite{CazalillaRigol,Polkovnikovetal}.

  Interestingly, it was first conjectured  by Rigol and
coworkers~\cite{Rigol_gge} that the steady state of simple few-body observables of
integrable systems following a quantum quench can be described by a generalized
Gibbs ensemble (GGE). The  density matrix of the GGE is obtained as
the less biased guess~\cite{Jaynes}  of the steady state given the 
constraints on the dynamics stemming from the existing set of non-trivial integrals of motion.
Surprisingly, it was found that, in order to reproduce
few body observables, only a subset of the order of $L$ (where $L$ is the system size)
of simple integrals of motion is needed~\cite{Rigol_gge}.

 Concerning the general applicability of the GGE,  there has been also some
debate~\cite{Silva} about the importance for thermalization of the locality
of the operators in the basis of eigenmodes of the system.  For the quantum
Ising chain, it was recently shown analytically that correlation functions of nonlocal operators
also thermalize to the GGE~\cite{Calabrese_Ising}.
Similar results had been found earlier for the Luttinger~\cite{Cazalilla}
and sine-Gordon models~\cite{Iucci2}.

 However, the reason why the GGE has been so successful in explaining the steady-state
correlations in very diverse models has remained rather obscure.  For local operators in
certain integrable field theories, solid arguments in favor of the validity of the GGE have been put
forward by  Fioretto and Mussardo~\cite{FiorettoMussardo}.
Furthermore, Cassidy and coworkers~\cite{Cassidy} recently introduced a generalization of the  eigenstate
thermalization hypothesis (ETH)~\cite{ETH,Rigol_ETH}
for integrable systems. Previously, the ETH has been successfully used to understand
thermalization in non-integrable systems~\cite{Rigol_ETH}.

 In this work, we describe a general method to demonstrate the applicability of the GGE
 in exactly solvable models for a general class of initial states. 
Our method does not require the explicit evaluation of correlation functions at asymptotically long times after the quench.
Instead, it suffices to show 
that the asymptotic correlation functions of  (either local or nonlocal) operators 
depend only on the expectation values of quasi-particle occupation operators
in the initial state. This property, together with certain properties of the class of initial states considered in
this work~\cite{MCChung,MingPeschel}, allows to demonstrate that each eigenmode of the system is subject to
a (mode) dependent effective temperature, the latter being nothing but a restatement of the GGE conjecture. 
This new point of view on the GGE  also explains some of the less well 
understood aspects of the  conjecture that have been briefly mentioned above. 
 As a matter of fact, it explains why the only set of integrals of motion that are needed
 to construct the GGE correspond to the quasi-particle occupation operators. For the class of exactly solvable
 models discussed below,  the latter  are a minimal set of $L$ integrals of motion that entirely determine
 the asypmtotic correlations.  The fact that the asymptotic correlation functions depend only on the expectation
value of these non-trivial integrals of motion means that the system remembers 
much more information about its initial conditions than it is the case in systems exhibiting thermalization to
a standard Gibbs ensemble.  For the latter, only the expectation value of the energy and the Hamiltonian
suffice to determine the effective temperature and chemical potential of the standard Gibbs ensemble 
describing thermal equilibrium of systems in the thermodynamic limit. 
The lack of relaxation of  correlation functions  to thermal equilibrium  found in this work 
bears a strong resemblance with the nonergodicity of the magnetization in the XY model
discussed several decades ago by McCoy~\cite{McCoyII}  and Mazur~\cite{Mazur} (see~\cite{Polkovnikovetal} 
for a recent review of this result). Thus, we shall call this behavior of the asymptotic correlations
 `nonergodic'. 

  Our goal in this article will be  to  illustrate our method to demonstrate the applicability of the 
 GGE by  applying it to several  models that have been previously analyzed 
 either analytically~\cite{Sengupta04,Cazalilla,Calabrese_Ising}
or numerically~\cite{Rigol_gge,Cassidy}:
The quantum Ising chain (cf. Sec.~\ref{sec:qi}), the Luttinger model (cf. Sec.~\ref{sec:LM}), and 
the lattice gas of hard-core bosons in one dimension or
quantum XX spin chain (cf. Sec.~\ref{sec:xxchain}).  In the former  two cases,
we consider a quench from an initial state that does not break the (lattice) translational
invariance and it is therefore conceptually simpler. 
In Sec.~\ref{sec:xxchain}, we  turn to a
the more involved case in which the initial state is not
translationally invariant but conserve the particle number. 
In our analysis, we shall consider correlations of both local and nonlocal operators,
demonstrating, for a broad class of initial states,
that in both cases thermalization to the GGE takes place. We also consider a more general
type of initial states that those that have been investigated in the past~\cite{Sengupta04,Cazalilla,Calabrese_Ising}.
We have relegated to the appendices the discussion of some of the most 
technical aspects of this work.

\section{The Quantum Ising chain}\label{sec:qi}

 Let us begin with an precise statement of the problem that we intend to address.
A quantum quench refers to the situation in which a system is prepared at $t =0$ in
an initial state (denoted $\rho_0$ below)  that is not an eigenstate of the
Hamiltonian $H$. Furthermore, we shall assume that, following the quench,
the system reached some kind of equilibrated state where  observables and 
correlation functions acquire (time-averaged) values about which they exhibit small temporal
fluctuations (cf. Fig.~\ref{fig:corrplot3}). A necessary condition for this equilibration to 
occur is that if we expand $\rho_0$ in the basis of eigenstates of $H$, 
\begin{equation}
\rho_0 = \sum_{n,m} C_{nm} |n \rangle \langle m |,
\end{equation}
the  coefficients $C_{mn}$  are sufficiently nonsparse on the basis of eigenstates of 
$H$ such that (for a sufficiently large system) unitary evolution can lead to a steady state 
as a result of dephasing between the contributions of many different eigenstates to the expectation
value of observables and correlation functions. The general conditions for equilibration have
been discussed in Ref.~\cite{Reimann}.

 We next begin our investigation of quantum quenches in exactly solvable models by considering 
the quantum Ising chain. For this system, the Hamiltonian that describes the time-evolution of the system
following the initial state preparation takes the form:
\begin{equation}
H = -J \sum_{j=1}^L \left[ \sigma^x_j \sigma^x_{j+1}  + g \, \sigma^z_j \right]\label{eq:hamqi}
\end{equation}
where $\sigma^x_j$ and $\sigma^z_j$ are the Pauli matrices at site $j$
and $J$ and $g$ are the model parameters.  As reviewed in Appendix~\ref{app:eigmodes},
the Hamiltonian in Eq.~(\ref{eq:hamqi}) can be diagonalized by means of a
non-local transformation due to Jordan and Wigner, which uncovers the
fact that its elementary excitations are indeed free fermions described by
\begin{equation}
H = \sum_{k} \epsilon_g(k) \left[  \gamma^{\dag}(k) \gamma(k) - \frac{1}{2} \right],
\end{equation}
where $\epsilon_g(k)  = 2J \sqrt{1+g^2 - 2 g \cos k}$ is the fermion dispersion ($|k| < \pi$).
The operators $\gamma(k)$ are the eigenmodes of the system for the set of parameters $(J,g)$.
They evolve according to the law: $\gamma(k,t) = e^{iHt/\hbar} \gamma(k) e^{-i H t/\hbar} =
e^{-i\epsilon_g(k)t/\hbar} \gamma(k)$.
However, the actual observables of the system correspond to the Pauli matrices, $\sigma^x_i$
and $\sigma^x_i$. For  example (cf. appendix~\ref{app:eigmodes}),
\begin{equation}
\sigma^z_j = 1-2 f^{\dag}_j f_j,
\end{equation}
where
\begin{align}
f_j(t)  &= \sum_k \left[ u_k(x)e^{-i \epsilon_g(k) t/\hbar}  \gamma(k) \right. \notag \\
 &\qquad \left. + v^*_k(x) e^{+i \epsilon_g(k) t/\hbar } \gamma^{\dag}(k) \right]
\end{align}
where $\tan \theta_g(k) =\sin k/(\cos k - g)$ and $u_k(x) = e^{i k x} \cos (\theta_g(k)/2)/\sqrt{L}$
and $v_k(x) = -i e^{ik x} \sin (\theta_g(k)/2)/\sqrt{L}$, such that  e.g. $|u_k(x)|^2 + |v_k(x)|^2 = L^{-1}$.

The class of initial states  with which we shall be
concerned in what follows is described
by a density matrix  $\rho_0 = Z^{-1}_0\, e^{-H_0/T}$,   where  $Z_0$ is a normalization constant
and  the operator $H_0$ is a quadratic form of the eigenmode
operators $\gamma(k)$ and $\gamma^{\dag}(k)$ (cf. Eq.~\ref{eq:initstate}). Since $H_0$ must be
hermitian, it  can be interpreted as
the Hamiltonian of the system at time $t \leq 0$, and the parameter
$T$ as the absolute temperature of an energy reservoir with which the system
was in contact (the pure state case is obtained by taking $T\to 0$).
The contact with the reservoir is removed at $t = 0$ as the Hamiltonian is
suddenly changed from $H_0$ to $H$ and the system allowed to evolve unitarily. This
defines the kind of quantum quench that has been analyzed in most cases so
far~\cite{Rigol_gge,Cazalilla,attention,Silva,Calabrese_Ising}.
Thus, the most general form for the initial
Hamiltonian $H_0$ reads:
\begin{multline}
H_0 = \sum_{k,k^{\prime}} \left[ \epsilon_0(k) \delta_{k,k^{\prime}} + V_0(k,k^{\prime})
\right]  \gamma^{\dag}(k) \gamma(k^{\prime}) \\
+ \sum_{k,k^{\prime}} \left[  \Delta^*_0(k,k) \gamma(k)\gamma(k^{\prime}) + \Delta_0(k,k^{\prime})
\gamma^{\dag}(k^{\prime}) \gamma^{\dag}(k) \right]. \label{eq:initstate}
\end{multline}
The term proportional to $V_0(k,k^{\prime})$ in Eq.~\eqref{eq:initstate} can
be interpreted as a scattering potential that is switched off at $t = 0$.
The presence of the scattering potential in $H_0$ means that, in general,
the initial state, $\rho_0$, breaks the  translational invariance of the lattice. Furthermore,
the last two terms in Eq.~\eqref{eq:initstate}
imply that that the number of fermion quasi-particles is not well defined in the initial state
 because $\left[\rho_0, N \right] \neq 0$,
where $N =  \sum_{k} \gamma^{\dag}(k)\gamma(k)$ is the quasi-particle number operator.

 The states introduced above
have two important properties:
 \emph{i}) Correlations of products of an arbitrary (even) number of
 Fermi operators like $\gamma(k)$, $\gamma^{\dag}(k)$, or
$f_i$ and $f^{\dag}_i$,
can be expressed in terms of products of correlation functions
of bilinear operators like $\langle \gamma^{\dag}(k) \gamma(k^{\prime})\rangle = \mathrm{Tr}\, \rho_0 \:
\gamma^{\dag}(k) \gamma(k^{\prime})$, $\langle \gamma(k) \gamma(k^{\prime})\rangle$, etc.
This result is known as Wick's (or more precisely,  Bloch-de Dominicis'~\cite{Bogolubov})
theorem and it is needed to show that the correlation functions of
nonlocal operators can be obtained from those of local operators (see below and Appendix~\ref{app:wick}); 
 \emph{ii}) for any partition of the eigemodes into two disjoint subsets $A$ (called ``system'' in what follows) and
$B$ (called ``environment''),  the reduced density matrix obtained by tracing out the
environment B  can be also written as the exponential of a quadratic form of the
Fermi operators $\gamma(k)$ and $\gamma^{\dag}(k)$~\cite{MCChung,MingPeschel}. In particular,
if the ``system'' A consists of a single eigenmode (and therefore $B$ contains
 the remaining $L-1$ modes), the reduced density matrix
 \begin{equation}
 \rho(k) = \mathrm{Tr}_{k^{\prime}\neq k} \: \rho_0
= Z^{-1}(k) e^{-\lambda(k) I(k)}, \label{eq:rdm}
\end{equation}
where the symbol $\mathrm{Tr}_{k^{\prime}\neq k}$
stands for the partial trace over all modes but $k$ and $I(k) = \gamma^{\dag}(k)\gamma(k)$ is the
quasi-particle occupation operator. This result applies to systems with
eigenmodes  obeying
both Fermi and Bose statistics~\cite{MCChung,MingPeschel}. 
It is worth noting that
the GGE density matrix is constructed as the direct product of these single-mode reduced density matrices:
\begin{equation}
 \rho_\mathrm{GGE} = \bigotimes_{k} \rho(k), \label{eq:ggerho}
\end{equation}
which is nothing but the mathematical statement that each mode is subject to 
a mode-dependent effective temperature $T(k) =  \lambda(k)/\epsilon(k)$. In fact, 
our method to prove the applicability of the GGE will rely on this interpretation of the
GGE. 

 In order to make contact with previous studies of the quantum Ising chain~\cite{Sengupta04,Silva,Calabrese_Ising},
we shall  analyze the following the case of an initial state that respects the lattice translation symmetry. This
requires that $V_0(k,k^{\prime}) = 0$ and  $\Delta_0(k,k^{\prime}) = \frac{i}{2} h_0(k) \delta_{k+k^{\prime},0}$. 
For specific choices of $\epsilon_0(k)$ and
$h_0(k)$ such that $\epsilon_0(k)/h_0(k) = \tan \phi(k)$, with $\phi(k) = \theta_g(k) - \theta_{g_0}(k)$, $H_0$
would correspond to the Hamiltonian of the quatum Ising chain at a different value of the parameter $g = g_0$.
At such point, $\epsilon_{g_0}(k)= \sqrt{\epsilon^2_0(k) + h^2_0(k)}$, is the dispersion of the quasi-particles
which are described by a different set of eigenmodes related to $\gamma(k)$ and $\gamma^{\dag}(k)$ by a canonical transformation parametrized by the angle $\phi(k)$~\cite{Sengupta04}. 
Thus, for such a choice we can speak of a quench in the parameter $g$.
However, for arbitrary $\epsilon_0(k)$ and $h_0(k)$, $H_0$ does not map to a quantum Ising chain Hamiltonian,
and therefore, our choice of the initial state of the quench, albeit translationally invariant,  is more general than
previous choices~\cite{Sengupta04,Silva,Calabrese_Ising}, which focused on quenching the parameter $g$ only.

 We next turn to the correlation functions of the model following the quench. We begin with the discussion
of the correlation function for a local operator such like the Fermi field:
\begin{align}
&C^{(2)}_{ff}(x_i-x_j, t) = \langle f^{\dag}_i(t) f_j(t) \rangle \\
 &\quad = \sum_{k,k^{\prime}} \left[ u^*_k(x_i) u_{k^{\prime}}(x_j) e^{i\left[\epsilon_g(k) - \epsilon_g(k^{\prime}) \right]t/\hbar}
 \langle \gamma^{\dag}(k) \gamma(k^{\prime} ) \rangle \right. \notag\\
&\left. \qquad + v_k(x_i) v^*_{k^{\prime}}(x_j) e^{-i\left[\epsilon_g(k) - \epsilon_g(k^{\prime}) \right]t/\hbar}
 \langle \gamma(k) \gamma^{\dag}(k^{\prime} ) \rangle \right] \notag\\
 &\quad  + \sum_{k,k^{\prime}} \left[ u^*_k(x_i) v^*_{k^{\prime}}(x_j) e^{i\left[\epsilon_g(k) + \epsilon_g(k^{\prime}) \right]t/\hbar}
 \langle \gamma^{\dag}(k) \gamma(k^{\prime} ) \rangle \right. \notag\\
&\left. \quad + v_k(x_i) u_{k^{\prime}}(x_j) e^{-i\left[\epsilon_g(k) + \epsilon_g(k^{\prime}) \right]t/\hbar}
 \langle \gamma(k) \gamma^{\dag}(k^{\prime} ) \rangle \right].
\end{align}
Thus, at any finite $t$ and for an arbitrary initial state, the above correlation function is fully determined by the
eigenmode correlations in the initial state $G_0(k,k^{\prime}) = \langle \gamma^{\dag}(k) \gamma(k^{\prime})\rangle$ and
$\tilde{F}_0(k,k^{\prime}) = \langle \gamma(k) \gamma(k^{\prime}) \rangle$. However,
the invariance of the initial state with respect to lattice translations greatly simplifies the above expression
implying that $G_0(k,k^{\prime}) = \delta_{k,k^{\prime}} N_0(k)$ and
$\tilde{F}_0(k,k^{\prime}) = F_0(k) \delta_{k+k^{\prime},0}$. Hence,
\begin{align}
&C^{(2)}_{ff}(x_i-x_j, t) = \sum_{k} \left[u^*_k(x_i) u_{k}(x_j) N_0(k)  \right.\notag \\
& \qquad  \left. + v_k(x_i) v^*_{k}(x_j) \left(1+ N_0(k)\right)   \right] \notag\\
&+ \sum_{k} \left[ u^*_k(x_i) v_{k}(x_j)  F_0(k) e^{2i\epsilon_g(k) t/\hbar} \right. \\
&\left.  + u_k(x_i) v^*_{k}(x_j) F^*_0(k) e^{-2i \epsilon_g(k) t/\hbar} \right] \label{eq:c2int}
\end{align}
We shall  next consider the limit $t\to +\infty$ of the above expression
after taking the thermodynamic limit where  $L\to +\infty$. Note that
$u^{*}_k(x_i) v_k(x_j) = -i L^{-1} e^{ik(x_i-x_j)} \sin (\theta_g(k)/2) \cos (\theta_g(k)/2)$ and 
$F_0(k)$, which is itself a function of $\epsilon_0(k)$ and $h_0(k)$  are assumed to be
well behaved, smooth functions of $k$. Therefore, it follows, by virtue of  the Riemann-Lebesgue lemma,
that the second term in the right hand side of Eq.~\eqref{eq:c2int}  vanishes in the $t \to +\infty$ limit. Thus,
\begin{align}
D^{(2)}_{ff}(x_i-x_j) &= \lim_{t\to +\infty} C^{(2)}_{ff}(x_i-x_j,t) \\
&=\sum_k   \left[u^*_k(x_i) u_{k}(x_j) N_0(k)  \right.\notag \\
& \qquad  \left. + v_k(x_i) v^*_{k}(x_j) \left(1 - N_0(k)\right)   \right], \label{eq:df}
\end{align}
where the thermodynamic limit is implicitly understood. Note 
that the above result, Eq.~\eqref{eq:df}, means that this correlation function depends only on 
 the expectation values of the $L$ integrals of motion $I(k) = \gamma^{\dag}(k) \gamma(k)$.  Indeed, 
 $D^{(2)}_{ff}(x_i-x_j)$ is a (weighted) sum of the expectation values,
$N_0(k) = \langle I(k) \rangle = \mathrm{Tr} \rho_0 \, I(k)$. Hence, 
for each term of the sum over $k$, we can use the second 
of the properties of the class of states $\rho_0$ described above, namely,
we can trace out all the modes $k^{\prime} \neq k$ and write $N_0(k) = 
\mathrm{Tr}\: \rho(k) I(k)$, where $\rho(k)$ is given in Eq.~\eqref{eq:rdm},
with $\lambda(k) = \ln \left[ (N_0(k)-1)/N_0(k)\right]$. This result obtained via a 
partial trace  amounts to the statement that each eigenmode is subject to a mode-dependent
effective temperature, which is equivalent to conjecturing that the asymptotic state is 
described by the GGE density matrix, Eq.~\eqref{eq:ggerho}.  This result also implies that
the $C_{ff}(x_i-x_j,t)$ will not relax its thermal equilibrium value, a behavior that we
call  'nonergodic'~\cite{Mazur,McCoyII}.

 Similar results can be obtained for the asymptotic limit of
other correlation functions of local operators such as
$A_j = f^{\dag}_j + f_j$ and $B_j  =  f^{\dag}_j - f_j$.
We merely state here the results:
\begin{align}
D^{(2)}_{AA}(x_i-x_j) &= \lim_{t\to +\infty} C^{(2)}_{AA}(x_i-x_j,t) \\
&= \lim_{t\to +\infty} \langle A_i(t) A_j(t) \rangle = \delta_{ij}  \label{eq:daa}\\
D^{(2)}_{BB}(x_i-x_j) &= \lim_{t\to +\infty} C^{(2)}_{BB}(x_i-x_j,t) \\
&= \lim_{t\to +\infty} \langle B_i(t) B_j(t) \rangle = -\delta_{ij} \label{eq:dbb} \\
D^{(2)}_{AB}(x_i-x_j) &= \lim_{t\to +\infty} C^{(2)}_{AB}(x_i-x_j,t) \\
&= \lim_{t\to +\infty} \langle A_i(t) B_j(t) \rangle \notag\\
&=\frac{1}{L} \sum_{k}
\left[ -e^{-ik (x_i - x_j)} e^{-i \theta_g(k)} N_0(k)  \right. \notag \\
 &\quad + e^{ik (x_i - x_j)} e^{i \theta_g(k)}  \big(1 - N_0(k)\big) \big]. \qquad \label{eq:dab}
\end{align}
Again we find that the asymptotic correlation functions are
nonergodic, as they only depend on $N_0(k)$.  

Using the above results, we are now in a position to discuss the correlations
of a nonlocal operator such as $\sigma^x_i$. Nonlocal means that
this operator does not reduce to a simple linear combination of the
eigenmode operators $\gamma(k)$ and $\gamma^{\dag}(k)$.
Indeed (cf. Appendix~\ref{app:eigmodes}),
\begin{equation}
\sigma^x_i = (f^{\dag}_i + f_i) \prod_{j < i} (1 - 2 f^{\dag}_j f_j),
\end{equation}
that is, $\sigma^x_i$ involves an infinite product of local operators (in this case $f_i$ and $f^{\dag}_i$).
As it is discussed in  the Appendices~\ref{app:wick} and \ref{app:sqt}, the two-point correlation
function of $\sigma^x_i$ can be expressed, by means of Wick's theorem, in terms
of a finite product of (equal time) correlation functions of the local operators $A_i$ and $B_i$.
The existence of the $t\to +\infty$ limit of those
correlation functions  (cf. Eqs. \ref{eq:daa}, \ref{eq:dbb}, \ref{eq:dab})
suffices to ensure the existence of the asymptotic correlation function (cf. Appendix~\ref{app:sqt}):
\begin{align}
D^{(2)}_{xx}(x_i-x_j) &= \lim_{t\to +\infty} C^{(2)}_{xx}(x_i-x_j,t) \\
 &= \lim_{t\to+\infty} \langle \sigma^{x}_i(t) \sigma^{x}_j(t) \rangle.
\end{align}
In this limit (a thermodynamically large system is  implicitly assumed),
the above correlation function reduces to a finite Toeplitz determinant (see Appendix~\ref{app:sqt}) 
which depends on $D^{(2)}_{AB}(x_k-x_l)$ (cf. Eq.~\eqref{eq:dab}) with $i \leq k, l \leq j$.

 Thus, we conclude that just as for the local correlations
discussed above, the nonlocal correlations are also nonergodic and are given by the GGE, which assumes
a mode-dependent effective temperature. As a corollary, it also follows that
only the $L$ occupation numbers $N_0(k) = \langle \gamma^{\dag}(k) \gamma(k) \rangle$ are needed to
determine the asymptotic  correlations of both local and nonlocal operators. Other integrals of motion different
from  $I(k) = \gamma^{\dag}(k) \gamma(k)$, such like e.g. the products $I(k_1) I(k_2)$, etc. do not play a role
in determining the asymptotic correlations and in the GGE.  The set of $L$ occupation numbers, $N_0(k)$, 
amounts to much less information than the full initial-state correlations,
which are determined  by both $N_0(k)$ and $F_0(k)$ ($3L$ real numbers, in total). However,
these $L$ occupation numbers amount by far to much more information that the expectation value of
the energy $\langle H \rangle = \mathrm{Tr}\,  \rho_0 \: H$ and the particle number $\langle N \rangle$,
which determine the effective temperature and chemical potential in the case of thermalization to the
standard (grand canonical) Gibbs ensemble.

 In this section we have focused on the quantum Ising model, which exhibits  fermionic quasi-particles. However,
 this is not a limitation to our methods, as shown in the following section, where we deal with the Luttinger model
exhibiting bosonic quasi-particles. We have also required that the initial state respect  lattice translational
invariance. As we show in Sec.~\ref{sec:xxchain}, this is again not a serious limitation to demonstrate the
applicability of the GGE.

\section{Quench in the Luttinger Model}\label{sec:LM}

 Let us next consider a quantum quench in the
Luttinger model (LM)~\cite{LiebMattis,Cazalilla}, which is a model exhibiting
bosonic quasi-particles. The initial state is assumed to be
of the form  $\rho_0 \propto e^{-H_0/T}$,  where
\begin{multline}
H_{0} =  \sum_{k\neq 0} \hbar | k| \{ v_0(k)\:  b^{\dag}(k) b(k) \\
-  \frac{1}{2}g_0(k)  \left[ b^{\dag}(k) b^{\dag}(-k) + b(k)  b(-k)) \right]\},\label{eq:hinit}
\end{multline}
 $v_0(k)$ and $g_0(k)$ being regular functions
 at $k = 0$. The operators $b(k)$ and $b^{\dag}(k)$ obey Bose statistics:
 $\left[ b(k), b^{\dag}(k^{\prime}) \right] = \delta_{k,k^{\prime}}$, commuting otherwise.
They are eigenmodes of the  Hamiltonian
\begin{equation}
H = \sum_{k\neq0} \hbar v(k) |k| \:
b^{\dag}(k) b(k),
\end{equation}
which dictates the evolution of the system for $t \geq 0$
and which we assume to describe an interacting version of the LM.
 Differently from the XX chain studied in the previous section,
 the eigenmodes of the LM are bosonic. In the initial state $\rho_0$, Eq.~\eqref{eq:hinit},
 the  number of bosonic modes is not well defined since
 \begin{equation}
 \big[ H_0,
 \sum_{k\neq 0} b^{\dag}(k) b(k) \big] \neq 0.
 \end{equation}
 However, $\rho_0$ does commute
 with the momentum operator $P = \sum_{k} k b^{\dag}(k) b(k)$, which implies that
 that $\rho_0$ is a translationally invariant state.

  We shall assume below that the `fundamental' fermions
 of the model~\cite{LiebMattis,Cazalilla} also diagonalize $H_0$.  This amounts
 to assuming that $H_0$ describes a non-interacting version of the LM~\cite{Cazalilla}.
Therefore, we can regard this situation as a quench from the non-interacting to the interacting LM,
with interactions $V = H - H_0$  being suddenly turned on at $t = 0$ as
the contact  with a bath at absolute temperature $T$ is also removed~\cite{Cazalilla}.
This allows us to determine the relation of the eigenmodes to the observables of the system.

 Physical operators in the LM can be expressed in terms of exponentials or derivatives of
(chiral) boson fields defined as follows:
\begin{align}
 \phi_{\alpha}(x,t)  &=  e^{iHt/\hbar} \phi_{\alpha}(x) e^{-iH t/\hbar}
  \notag\\
 &= \phi_{0\alpha}(t) +  \frac{2\pi x }{L} N_{\alpha}  + \Phi_{\alpha}(x,t) + \Phi^{\dag}_{\alpha}(x,t),
 \end{align}
 where  $L$  is  the system size and  $N_{\alpha}$  is the number of fermions
 of chirality $\alpha = r, l$ ($\left[N_{\alpha}, \phi_{0\beta}\right] =  i\delta_{\alpha\beta}$).
 Below, we shall work in the sector where $N_r = N_l = 0$ which also contains
 the ground state of $H$, namely $|0\rangle$ (\emph{i.e.} $b(k) |0\rangle = 0$ for all $k$).
 Furthermore, in terms of the  eigenmodes~\cite{LiebMattis,Cazalilla},
\begin{multline}
\Phi_{\alpha}(x,t) =
  \sum_{k > 0} \left( \frac{2\pi}{k L}\right)^{\frac{1}{2}} e^{i s_{\alpha} k x} \: \big[ \cosh \theta(k) e^{-i v(k)|k| t} \\
 \times b(s_{\alpha} k)\ - e^{i v(k) |k| t} \sinh \theta(k) \: b^{\dag} (-s_{\alpha} k) \big],
 \end{multline}
 with  $s_{r} = -s_{l} =1$ and  $\tanh \theta(k) = g_0(k)/v_0(k)$.
 Using these chiral fields,  the density (or `current') operator for each
 fermion chirality reads
 \begin{equation}
 J_{\alpha}(x,t) = \, \, :\psi^{\dag}_{\alpha}(x,t) \psi_{\alpha}(x,t): \,\, = \frac{1}{2\pi} \partial_{x} \label{eq:jj}
 \phi_{\alpha}(x,t),
 \end{equation}
 $:\ldots:$ meaning normal order with respect to the ground state of $H_0$~\cite{LiebMattis}.
 Note that the $J_{\alpha}(x,t)$ are \emph{local}  in the eigenmodes, $b(k)$ and $b^{\dag}(k)$.
 On the other hand, the `fundamental' fermion fields~\cite{LiebMattis},
\begin{equation}
\psi_{\alpha}(x,t) \sim e^{i s_{\alpha} \phi_{\alpha}(x,t)} \label{eq:bosonization}
\end{equation}
are \emph{nonlocal} (`vertex')  operators. Using  Wick's theorem, we
can recast any fermion correlation function
in terms of two body correlators of the local fields $\phi_{\alpha}(x,t)$
because the cumulant expansion to second order is exact  for states like
$\rho_0$. Mathematically,
\begin{equation}
\langle e^{i A_{\alpha}(x_1, \ldots, x_n,t)} \rangle =
e^{-\frac{1}{2} \langle A^2_{\alpha}(x_1,\ldots, x_n,t) \rangle}, \label{eq:cexp}
\end{equation}
where ($\sum_{i} p_i =0$)
\begin{equation}
A_{\alpha}(x_1, \ldots, x_n,t) = \sum_{i=1}^{n} p_i \phi_{\alpha}(x_i,t).
\end{equation}
Eq.~\eqref{eq:cexp} can be proven by expanding in series the exponential
in the left hand side in a Taylor series and applying Wick's theorem to all the terms,
which involve powers of $A(x_1,\ldots, x_n)$. Resuming the resulting
series, the right hand side of \eqref{eq:cexp} is obtained.

 From the previous discussion, it can be seen that in order to compute the equal
 time correlations of the LM it is sufficient to consider
\begin{equation}
C^{(2)}_{\phi_r}(x,t) =   \langle \phi_r(x,t) \phi(0,t) \rangle =
 D^{(2)}_{\phi_r}(x) + F^{(2)}_{\phi_r}(x,t),
\end{equation}
where
\begin{multline}
D^{(2)}_{\phi_r}(x)  = \sum_{k \neq 0} \left(\frac{\pi}{|k| L} \right)   \left( \cosh 2\theta(k) + \mathrm{sgn}(k) \right) \\
\times \big[  e^{i k x}
\left(1 + N_0(k)\right)  + e^{-i k x} N_0(k) \big], \label{eq:m1}
\end{multline}
is the contribution of the diagonal correlations $N_0(k) = \langle b^{\dag}(k) b(k) \rangle$. However,
\begin{multline}
F^{(2)}_{\phi_r}(x,t) = \sum_{k\neq 0}  \left(\frac{\pi}{|k| L} \right)  \left( \cosh 2 \theta(k) + \mathrm{sgn}(k) \right)  \\
\times \left[ e^{ik x- 2 i v(k) |k| t} F_0(k)
+ e^{-i k x + i v(k) |k| t} F^*_0(k) \right], \label{eq:f2}
\end{multline}
where $F_0(k) =\langle b(k) b(-k) \rangle$, stems from the `anomalous' (\emph{i.e.} `superfluid') correlations.
Note that the translational  invariance of the initial state implies
that  $N_0(k)$ and $F_{0}(k)$ are the only non-vanishing
two-point correlations of the eigenmodes in the initial state.
Whereas the contribution of the diagonal correlations $N_0(k)$ is time independent,
the contribution of the anomalous terms depends on time. It may be expected that,  because of dephasing
between the different Fourier components (\emph{i.e.} the Riemann-Lebesgue lemma),   in the thermodynamic limit
$F^{(2)}_{\phi_r}(x,t)$ vanishes as $t\to +\infty$.
However, the  $t\to +\infty$ limit of this function limit must be handled with care because the
$1/|k|$  in Eq.~(\ref{eq:f2}) yields terms
diverging logarithmically as $t\to +\infty$~\cite{Cazalilla}.
Fortunately,  as we have described above (cf. Eqs.\ref{eq:jj}, \ref{eq:cexp}), only the derivatives or exponentials of
$C^{(2)}_{\phi_r}(x,t)$  appear  in  the physical correlation functions of the LM model.
For example, using \eqref{eq:bosonization}, the two-point correlation function of the
right moving Fermi fields reads:
\begin{equation}
\langle \psi^{\dag}_r(x,t) \psi_r(0,t) \rangle = A \exp\left[ C^{(2)}_{\phi_r}(x,t) - C^{(2)}_{\phi_r}(0,t) \right].
\end{equation}
Since (for $L\to +\infty$ and $T = 0$)
\begin{equation}
F_{\phi_r}(x,t) - F_{\phi_r}(0,t) \sim \log \left| \frac{(2v t)^2-x^2}{(2v t)^2}  \right|,
\end{equation}
where $v = v(k = 0)$, the time-dependent logarithmic contributions  disappear
(for finite $x$) as  $t\to +\infty$ ~\cite{Cazalilla}. Therefore, we can safely ignore
the contribution of $F^{(2)}_{\phi_r}(x,t)$ in the $t\to +\infty$ limit.
This means that all correlations are asymptotically determined by
$D_{\phi_r}(x)$, which only depends on   $N_0(k) = \langle b^{\dag}(k) b(k)\rangle$, \emph{i.e.}
it is nonergodic.  Furthermore, for each term of the sum in Eq.~\eqref{eq:m1}, we can trace
out all the $k^{\prime} \neq k$  and write
$N_0(k) = \text{Tr} \left[ \rho(k) b^{\dag}(k) b(k) \right] = \text{Tr}\: \rho_{\mathrm{GGE}}\: b^{\dag}(k)b(k)$.
Since  $\rho(k)  = \mathrm{Tr}_{k\neq k^{\prime}}\: \rho_0
 = Z^{-1}(k) \: e^{-\lambda(k) b^{\dag}(k)b(k)}$~\cite{MCChung}, we arrive
 at the same result as if we had used the GGE density matrix $\rho_{\mathrm{GGE}} = \bigotimes_{k}\rho(k)$.
Thus the equivalence with the GGE is established for the simple correlation funcitions
involving the bose field $\phi_{\alpha}(x)$ in the LM.

 Finally, it is interesting to note that  translationally invariance requires that eigenmode
 correlations are bi-partite, that is,
each mode $k$ is correlated only with the eigenmode at $-k$ (cf. Eq.~\eqref{eq:hinit}).
Thus, an alternative way of obtaining the  results of this section and those of Sec.~\ref{sec:qi} is
 to compute the reduced density $\rho(k)$ as a partial trace for a partition of the
eigenmodes into $k > 0$ and $k < 0$. Thus, we can regard the effective temperature
$T(k) = \lambda(k)/v(k)|k|$ for e.g. the modes with $k > 0$  as the result of their correlations
with the $k < 0$ modes (and viceversa)~\cite{MCChung}.

\section{The XX chain} \label{sec:xxchain}

The Hamiltonian of  the XX chain reads
\begin{equation}
H = -\frac{J}{2}\sum_{i=1}^L  \left[ \sigma^{+}_{i}
\sigma^{-}_{i+1} + \sigma^{-}_i\sigma^{+}_{i+1}\right ] + \frac{h}{2}\sum_{i=1}^L \sigma^z_i, \label{eq:hamxx}
\end{equation}
in terms of the Pauli  matrices $\sigma^{\pm}_i, \sigma^z_i$.
We shall assume an open chain like in Ref.~\cite{Rigol_gge}.
In order to diagonalize the Hamiltonian, we first carry a Jordan-Wigner transformation
to express the Pauli matrices in terms of Fermi operators $f_i, f^{\dag}_i$ and Fourier expand
the latter  in terms of $f(k)$ and $f^{\dag}(k)$ (cf. Appendix~\ref{app:eigmodes}). Hence,
\begin{equation}
H = \sum_{k} \epsilon(k) f^{\dag}(k) f(k), \label{eq:hamxxd}
\end{equation}
with $\epsilon(k) = - J \cos k a_0 -h$ ($0< k < \pi$). Thus, the eigenmodes
of the system are described by the Fermi operators $f(k)$ and $f^{\dag}(k)$, which evolve
in time according to $f(k,t) = e^{i H t/\hbar} f(k) e^{-i Ht/\hbar}
= e^{-i \epsilon(k) t/\hbar} f(k)$, etc.

The initial state is  given by the density matrix $\rho_0 = Z^{-1}_0 \: e^{-H_0/T}$, where
\begin{equation}
H_0 = \sum_{k,k^{\prime}} \left[ \epsilon_0(k) \delta_{k,k^{\prime}} + V_0(k,k^{\prime})
\right]  f^{\dag}(k) f(k^{\prime}).  \label{eq:initstate2}
\end{equation}
 In order to make contact with the numerical studies of Ref.~\cite{Rigol_gge},
we have assumed that the initial state commutes with $N$ in Eq.~\eqref{eq:initstate2}.
Therefore, the anomalous terms (such like those $\propto \Delta_0(k,k^{\prime})$ in Eq.~\ref{eq:initstate})
are absent in this case. However, the presence of the scattering potential $V_0(k,k^{\prime})$
implies that the initial state $\rho_0$ breaks the lattice translational invariance.

We next turn to the analysis of correlation functions. We first consider the equal time
correlation of a  \emph{local} operator like $O(x_i) = \sum_{k} \varphi_k(x_i)\: f(k)$,
where  we shall require that (the square of) $\varphi_k(x_i)$ is normalized to the system
size (\emph{i.e.} $|\varphi_k(x_i)| \sim O(L^{-1/2})$). This means that  the quasi-particles of the system
(described by the eigenmodes $f(k)$ and $f^{\dag}(k)$ of the Hamiltonian $H$) can occupy  extended
orbitals after the quench and are not localized. In other words, in the thermodynamic limit
the quasi-particle spectrum of $H$ is assumed to be described by  a continuum of extended (i.e spatially delocalized)
levels with no macroscopic degeneracies. In principle, violation of this requirement 
may prevent the system from reaching equilibration as contributions from localized 
states will lead to oscillatory  behavior at
long times and the absence of decoherence. With this caveat, let us consider:
\begin{multline}
C^{(2)}_O(x_i,x_j,t) = \langle O^{\dag}(x_i,t) O(x_j,t) \rangle\\ = \sum_{k,k^{\prime}} \varphi^*_k(x_i) \varphi_{k^{\prime}}(x_j) \:
 G_0(k,k^{\prime}) e^{-i\left[ \epsilon(k) - \epsilon(k^{\prime}) \right]t/\hbar},\label{eq:cfint}
\end{multline}
which depends on the eigenmode correlations $G_0(k,k^{\prime}) =  \langle f^{\dag}(k) f(k^{\prime}) \rangle$.
The latter  are $L^2$ real  numbers containing the full information about the initial state~\cite{MCChung,MingPeschel}.  
With the above assumptions and in the thermodynamic limit, we find that (see discussion below)
\begin{align}
D^{(2)}_{O}(x_i,x_j) &= \lim_{t\to +\infty} C^{(2)}_O(x_i,x_j,t) \notag\\
 &= \sum_{k}\varphi^*_k(x_i)\varphi_{k}(x_j) 
N_0(k), \label{eq:coinf}
\end{align}
where $N_0(k) = \langle I(k)\rangle$ with $I(k) = \gamma^{\dag}(k)\gamma(k)$ 
is the quasi-particle occupation in the initial state.
Eq.~\eqref{eq:coinf} means that $D^{(2)}_O(x_i,x_j)$ is nonergodic~\cite{McCoyII,Polkovnikovetal} as it
is entirely determined by $L$ real numbers, the quasi-particle occupations
$N_0(k) = \mathrm{Tr} \left[ \rho(k) I(k) \right] )$,
where $\rho(k)$ has been defined above (cf. Eq.~\ref{eq:rdm}).  Hence, we can again
perform the partial trace in each of the terms of the sum \eqref{eq:coinf}, and conclude that
each eigenmode is subject to a $k$-dependent effective temperature, as expected from the GGE.
This result also implies that the asymptotic correlation functions are determined by much less 
information than the one contained in the initial state (\emph{i.e.} $O(L)$ vs. $O(L^2)$ real numbers).
Yet, this is much more information than the one needed to characterize the asymptotic state
of standard thermal equilibrium.

\begin{figure}
\includegraphics[width=0.5\textwidth]{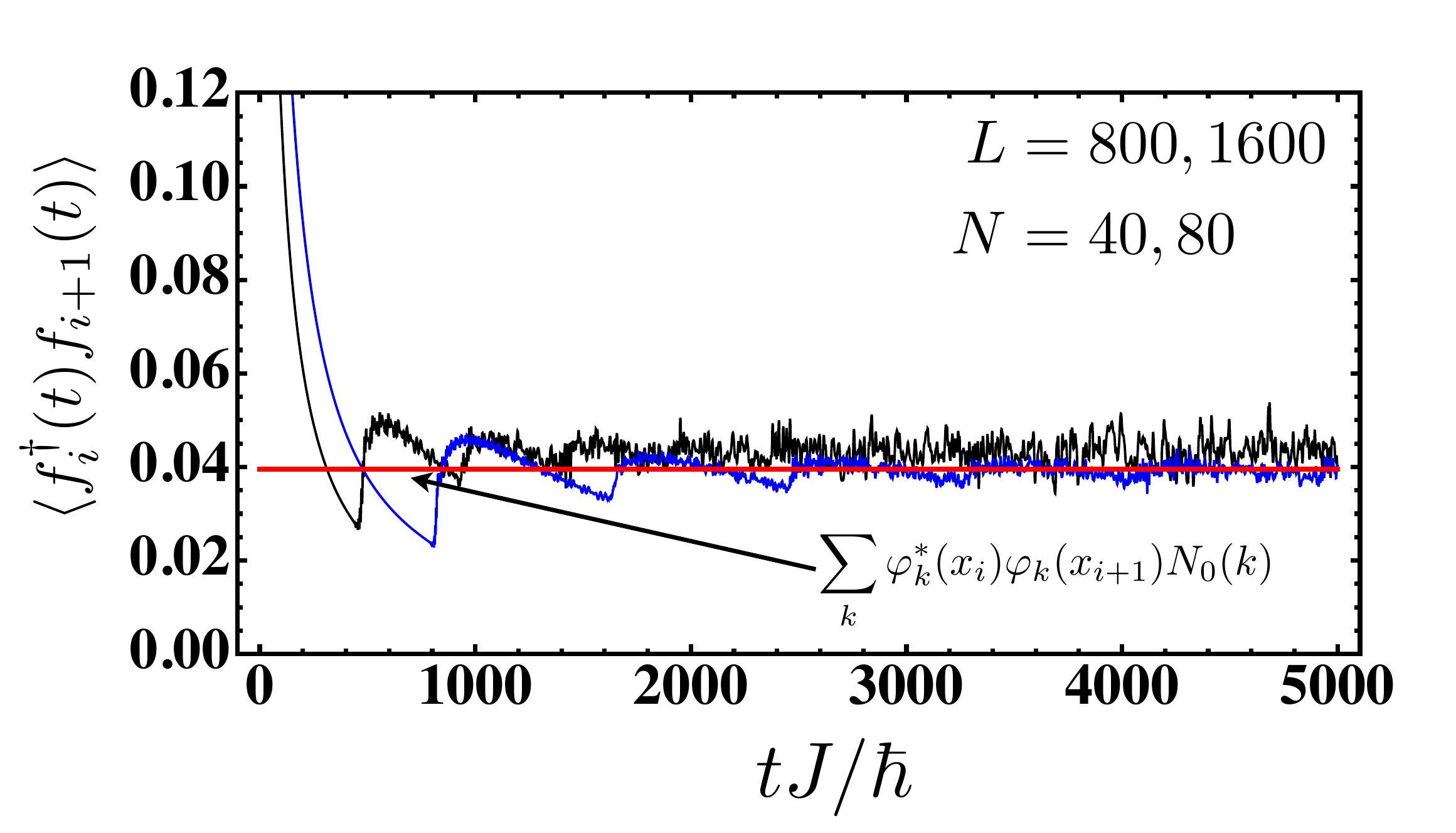}
\caption{ (color online)
Time evolution of the correlation function of the local operator $O(x_i) = f_i$ for $i=L/2$ and $L/2+1$,
$\langle f^{\dag}_i(t) f_{i+1}(t) \rangle$,  for $L=800$ and $L=1600$, with  $N=40$ and $N=80$ . The horizontal
line corresponds to the time average for the largest system size ($L=1600$ and $N=80$).
In the initial state $N=40$ ($N=80$) hardcore bosons that are confined  by a harmonic potential of the form 
$V_0(x_i-x_M)^2$, with $V_0 = 10^{-3} \:J/\sqrt{N}$ and $x_M = L/2$. 
\label{fig:corrplot3}}
\end{figure}
\begin{figure}
\includegraphics[width=0.5\textwidth]{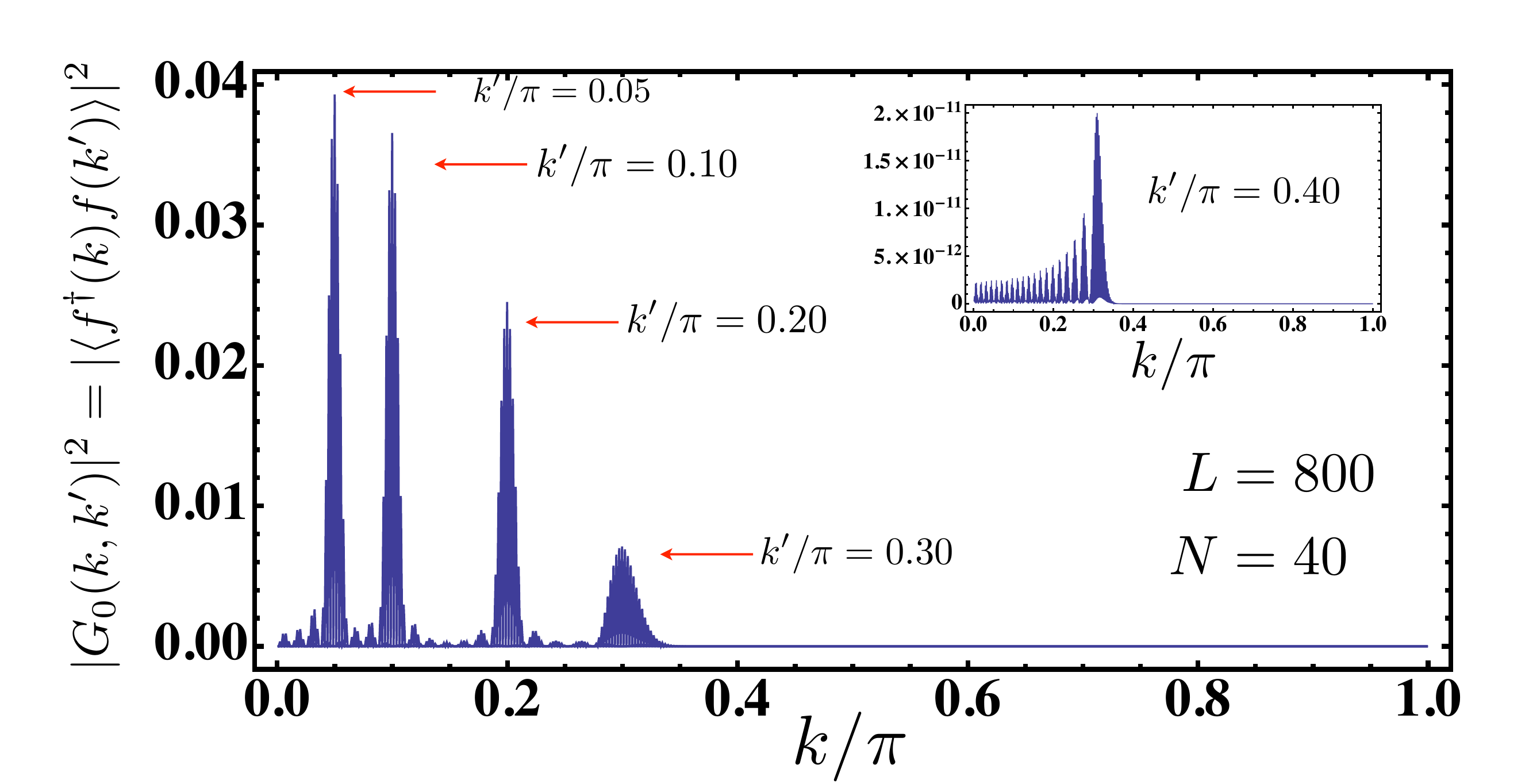}
\caption{ (color online)
Modulus square of the eigenmode correlations, $|G_0(k,k^{\prime})|^2 = |\langle f^{\dag}(k) f(k^{\prime}) \rangle|^2$ in the initial state of a XX chain at $T=0$,  for a  system in a box of size $L=800$ containing $N = 40$ hardcore bosons, and $k^{\prime}/\pi = 0.05, 0.1, 0.2, 0.3$ and $0.4$ (inset). The initial state describes  $N=40$ hardcore bosons that are confined  by a harmonic potential with the same parameters as in Fig.~\ref{fig:corrplot3}. Note that the eigenmode correlations $|G_0(k,k^{\prime})|^2$ are strongly peaked at $k =  k^{\prime}$. For $k > k_0$ with $k_0/\pi \simeq 0.35$ for $N= 40$ (see inset) the correlations are no longer peaked at $k\simeq k^{\prime}$. However, they become substantially smaller than the typical peak values at $k^{\prime} < k_0$ ($k_0$ increases with $N$, cf. Fig.~\ref{fig:corrplot4}).
\label{fig:corrplot1}}
\end{figure}
\begin{figure}
\includegraphics[width=0.5\textwidth]{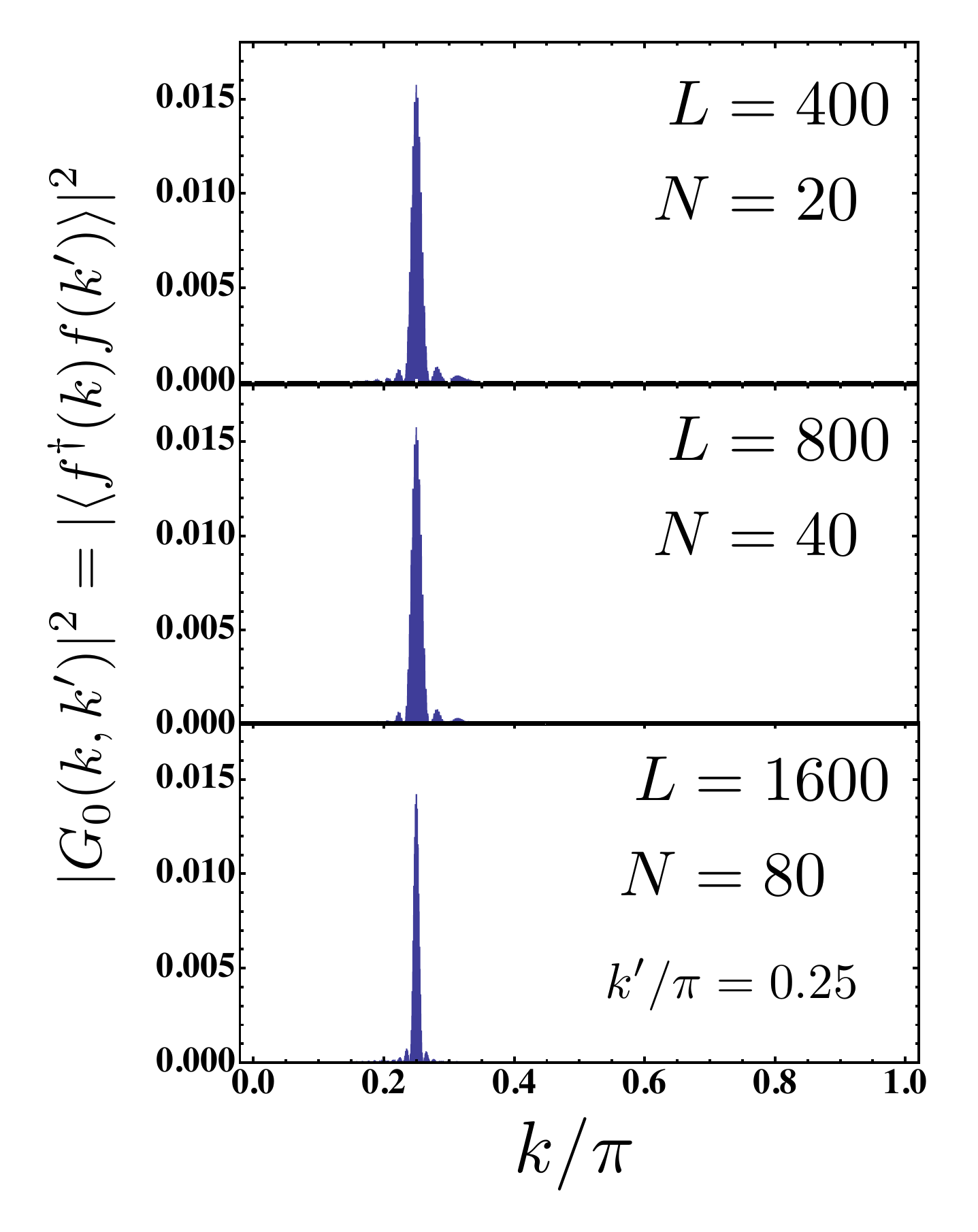}
\caption{ (color online)
Finite-size scaling of the Modulus square of the eigenmode correlations, $|G_0(k,k^{\prime})|^2 = |\langle f^{\dag}(k) f(k^{\prime}) \rangle|^2$ for the same initial state of a XX chain as described in the caption of Figs.~\ref{fig:corrplot3} and \ref{fig:corrplot1}  for  systems of size $L=400, 800, 1600$ and $k^{\prime}/\pi = 0.25$. Note that the peak becomes narrower as the system approaches the thermodynamic limit where $N$ and $L \to +\infty$ while $n_0 = N/L = $ const.
\label{fig:corrplot2}}
\end{figure}

  In order to demonstrate Eq.~\eqref{eq:coinf}, we display in Fig.~\ref{fig:corrplot3}.
 the results of the numerical evaluation of the time evolution of  $C_{ff}^{(2)}(x_i,x_{i+1},t)$ ($i = L/2$) using 
 Eq.~\eqref{eq:cfint} for $O(x_i) = f_i$.  We consider  an initial state $\rho_0$ for which  $V_0(k,k^{\prime})$ in
Eq.~\eqref{eq:initstate2} is a harmonic potential that confines $N$ hardcore bosons at the center
of an open chain of  $L$ sites. The potential $V_0(k,k^{\prime})$ is taken to scale as $ V_0/\sqrt{N}$ ($V_0 = 10^{-3}\, J$ 
in Figs.~\ref{fig:corrplot3},\ref{fig:corrplot1}, and \ref{fig:corrplot2}) in order to obtain
a well defined thermodynamic limit of the initial cloud of hardcore bosons~\cite{Cassidy,RigolMuramatsu}.
This potential is switched off at $t = 0$
and the  bosons are allowed to expand~\cite{Rigol_gge}.
The vertical line in this figure corresponds to the time
average  (for $L = 1600$, the average for  $L=800$  is not shown but it is very close to it). 
The average is given by Eq.~(\ref{eq:important}) evaluated at finite $L$. It can be seen from Fig.~\ref{fig:corrplot3} 
that for both $L=800$ and $L  = 1600$, after a short transient, the correlation function exhibits roughly equilibrationa
and its time fluctuations about the average  become fairly small. 
In can be also seen that, as $L$ increases from $800$ to $1600$ (while keeping $N/L$ constant), 
the size of the time fluctuations decreases suggesting that for $L\to +\infty$ they will vanish. Therefore,  in the thermodynamic limit the asymptotic correlations are given by the quasi-particle occupation $N_0(k)$. In the Appendix~\ref{app:wick}, we further explore the equivalence between the thermodynamic limit of time-averaged correlations and their $t\to+\infty$ limit after taking the
thermodynamic limit.

 To understand the  behavior displayed in Fig.~\ref{fig:corrplot3} in physical terms, note that,
in the thermodynamic limit, the sums  over $k$ and $k^{\prime}$ in the expression for $C^{(2)}_O(x_i,x_j,t)$  
become integrals and dephasing between different eigenmode contributions to Eq.~\eqref{eq:cfint}  leads to 
the decay  in time  of the correlations except  for the terms where $\epsilon(k) = \epsilon(k^{\prime})$. 
It is worth investigating how this dephasing takes place in  more detail 
because, generally speaking, for non-translationally invariant
states,  $G_0(k,k^{\prime})$ is not generally speaking a  
smooth function of $k$ and $k^{\prime}$ (cf. figure~\ref{fig:corrplot1} and \ref{fig:corrplot2}). Thus, arguments
based on the Riemann-Lebesgue lemma similar to those employed in sections~\ref{sec:qi} and \ref{sec:LM} for 
translationally invariant states are not applicable. However, 
for non-translationally invariant states (\emph{i.e.} $V_0(k,k^{\prime})\neq 0$), we  numerically 
find that, as the thermodynamic limit is approached (cf. Fig.~\ref{fig:corrplot2})
\begin{equation}
|G_0(k,k^{\prime})|^2 \to N^2_0(k) \delta_{k,k^{\prime}} + \Delta R_0(k,k^{\prime}), \label{eq:important}
\end{equation}
where $\Delta R_0(k,k^{\prime})$ decays rapidly for $|k-k^{\prime}| \gg L^{-1}$. 
Eq.\eqref{eq:important} must be understood as the statement that typical correlations become strongly
peaked at $k = k^{\prime}$ as $L \to+\infty$. Thus, setting $G_0(k,k^{\prime}) \simeq N_0(k) \delta_{k,k^{\prime}}$
in Eq.~\eqref{eq:cfint} becomes an increasingly good approximation at large $t$ where decoherence
acts most efficiently on the contributions to  the double sum~\eqref{eq:cfint}  of   quasi-particle levels  $k$ and $k^{\prime}$ 
that are close in energy  and  correlated (i.e. for which $G_0(k,k^{\prime})$ is not negligibly 
small). 

\begin{figure}
\includegraphics[width=0.45\textwidth]{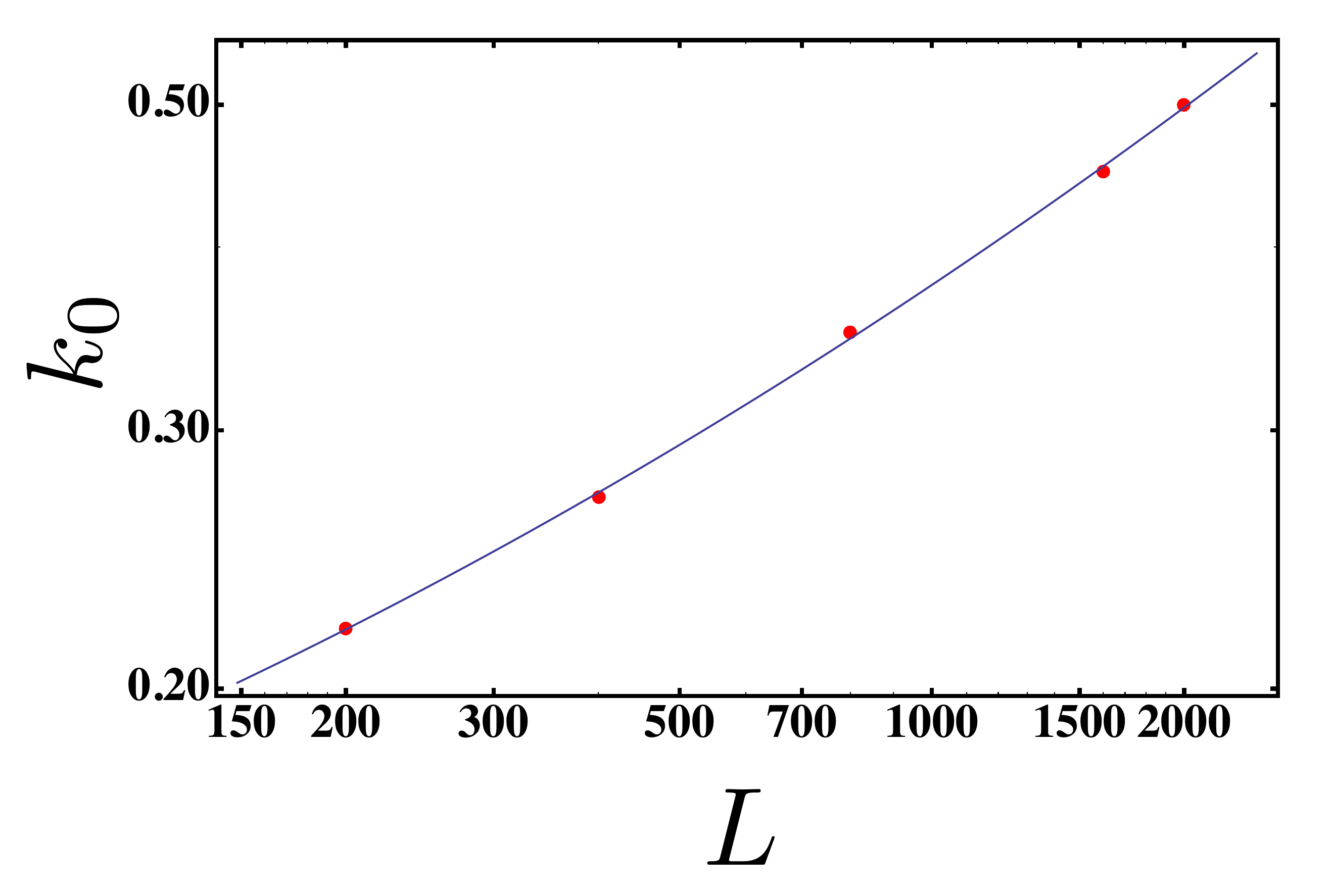}
\caption{ (color online)
Log-log plot of the size dependence of the cut-off $k_0$ for $n_0 = N/L = 0.05$. The continuous curve is just a gide
to the eye. Other parameters are the same used to generate figures~\ref{fig:corrplot3}, \ref{fig:corrplot1},  and \ref{fig:corrplot2}.
\label{fig:corrplot4}}
\end{figure}

    The claim of Eq.~\eqref{eq:important} is illustrated in figures~\ref{fig:corrplot1} and \ref{fig:corrplot2}
for the same system used to generate Fig.~\ref{fig:corrplot3}.  Fig.~\ref{fig:corrplot1} 
displays $|G_0(k,k^{\prime})|^2$ as a function of $k$
for several values of $k^{\prime}$, for a system of $L = 800$ sites and $N = \sum_k N_0(k) = 40$ hardcore bosons
at $T = 0$. It can be seen that $|G_0(k,k^{\prime})|^2$ is strongly peaked at $k = k^{\prime}$
for $k^{\prime} <  k_0$ ($k_0/\pi \simeq 0.35$ for $N = 40$). However, for $k^{\prime} > k_0$ (see inset)
the peak is no longer at $k=k^{\prime}$.  In this case, however, the values of $|G_0(k,k^{\prime})|^2$ become very small 
compared to  typical the peak values of $|G_0(k,k^{\prime})|^2$ 
 for $k < k_0$. The cut-off $k_0$ is determined by the number of hardcore
bosons in initial state $N \propto L$ (see discussion below and Fig.~\ref{fig:corrplot4}).  
As the system size $L$ increases while  keeping the
 lattice filling $n_0 =  N/L$ constant and scaling the initial trap strength as $V_0/\sqrt{N}$~\cite{RigolMuramatsu},
 the peak  of $G_0(k,k^{\prime})$ at $k = k^{\prime}$ becomes substantially narrower (cf. Fig.~\ref{fig:corrplot2}).
 We also found a similar behavior of $G_0(k,k^{\prime})$  when the potential that initially acts upon the hardcore bosons
 was taken to be an extended superlattice~\cite{unpub} and an infinite square well box of size smaller than $L$~\cite{unpub2}.  
 Although we cannot find a rigorous mathematical proof that Eq.~\eqref{eq:important} holds for any non-translationally
 invariant initial state of the form $\rho_0 \sim e^{-H_0/T}$, with $H_0$ given by  \eqref{eq:initstate2}. 
 we expect it to hold for any physically sensible  scattering potential $V_0(k,k^{\prime})$.
The following physical argument can be used to support this expectation.  
Recalling the relationship between the eigenmodes of $H$
and the eigenmodes of the initial Hamiltonian, $H_0$,
the two sets of operators are related by the canonical transformation, $f(k) = \sum_{\alpha} \phi_{\alpha}(k) f(\alpha)$;
$\phi_{\alpha}(k)$ are the eigenfunctions of $H_0$ in the basis of eigen orbitals of  $H$,
that is, $\phi_{\alpha}(k) = \langle k| \alpha\rangle$ where $H_0 |\alpha \rangle = (\epsilon_{\alpha} -\mu) |\alpha\rangle$.
Hence,
\begin{equation}
G_0(k,k^{\prime}) = \sum_{\alpha=1}^{L} \phi^*_{\alpha}(k)
\phi_{\alpha}(k^{\prime}) F(\epsilon_{\alpha},T), \label{eq:sumalpha}
\end{equation}
where $F(\epsilon_{\alpha},T) =
\left[ e^{(\epsilon_{\alpha}-\mu)/T} + 1\right]^{-1}$ are the thermal occupations
of the orbitals $\phi_{\alpha}(k)$ in the initial state and $\mu$ is the chemical potential. The latter 
allows us to fix the average number of quasi-particles in the initial state. Next, we notice that,
as the thermodynamic limit is approached (\emph{i.e.} $L\to +\infty$
while keeping  the lattice filling $n_0 =  N/L$ constant),
 Eq.~\eqref{eq:sumalpha} becomes an infinite sum. Correlations
are maximized for $k = k^{\prime}$ because the summands, $|\langle \alpha | k \rangle|^2$ 
are all positive and correspond to the probability that a quasi-particle   initially  found 
in the state $|\alpha\rangle$ ends up in the state $|k\rangle$
after the quench.  On the other hand, for $|k-k^{\prime}|\gg L^{-1}$, the eigenmodes become largely uncorrelated
because the sum in Eq.~\eqref{eq:sumalpha}
involves  a large number of complex  amplitudes
$\langle k| \alpha\rangle \langle\alpha| k^{\prime} \rangle =
\phi^*_{\alpha}(k) \phi_{\alpha}(k^{\prime})$, describing the quantum interference
between transitions where the quasi-particle ends in one of  two orthogonal orbitals, either $|k\rangle$ or
in $|k^{\prime}\rangle$.  As $L\to +\infty$, the amplitudes interfere
destructively and typically average to zero.  Furthermore, at low temperatures,  
the sum ~\eqref{eq:sumalpha} contains a cut-off which is roughly given 
by the state $\alpha_{\max}$ for which $F(\epsilon_{\alpha_{\max}}-\mu) \ll 1$.
Thus, for example, for $T=0$, $|\alpha_{\max}\rangle$ corresponds to the orbital with the $N$-th 
smallest eigenvalue  $\epsilon_{\alpha}-\mu$. As $k = k^{\prime}$ increases,  
the number of nodes of the orbital $|k\rangle$ in the
initially (harmonically) confined region increases, and the overlaps $\phi_{\alpha}(k) = \langle k | \alpha\rangle$  
dramatically decrease in magnitude.
This explains the existence of the cut-off $k_0$ seen in Fig.~\ref{fig:corrplot1}. Consistent with this effect,
we numerically observe (see Fig.~\ref{fig:corrplot4}) that $k_0$ increases as the number of hardcore bosons 
$N$ increases, which makes $\alpha_{\max}$ bigger. 
At higher temperatures, the increase in entropy implies that 
quasi-particles are spread even more over the set of initial orbitals $|\alpha\rangle$, and the 
quantum interference effects will be weakened thus decreasing the correlations 
$|G_0(k,k^{\prime}\neq k)|^2$ even further.

  To end this section, we shall use the above results to sketch the proof
 that the momentum distribution function of the hardcore bosons, $S(q,t)$, for $t \to +\infty$ is also nonergodic. 
 We first recall that
\begin{align}
S(q,t)  &= \frac{1}{L}   \sum_{ij}  e^{i q (x_i-x_j)}\: \langle \sigma^{-}_i(t) \sigma^{+}_j(t)  \rangle \\
&= \frac{2}{L}
\sum_{ij}  e^{i q (x_i-x_j)}\:  C^{(2)}_{xx}(x_i,x_j,t),
\end{align}
where $C^{(2)}_{xx}(x_i,x_j,t) = \langle \sigma^{x}_i(t) \sigma^{x}_j(t) \rangle$. The latter
correlation function involves the operator $\sigma^x$ that is nonlocal in the eigenmodes $f(k),f^{\dag}(k)$.
However,  using Wick's theorem, it can
be written in terms of products of two-point correlation functions
of the local operators $A_i = f^{\dag}_i + f_i$ and $B_i = f^{\dag}_i - f_i$.
The $t\to +\infty$ limit of the product exists provided the
$t \to +\infty$ limit  of the two-point correlation functions involved in the product also exists
(see the extended discussion in Appendix~\ref{app:wick}). Moreover,  as $t\to +\infty$
\begin{align}
\langle A_i(t) A_j(t) \rangle &\to \delta_{ij},\\
\langle B_i(t) B_j(t) \rangle &\to -\delta_{ij}.
\end{align}
Hence,  the $t\to +\infty$ limit of $C^{(2)}_{xx}(x_i,x_j,t)$ reduces to a Toeplitz determinant
 involving  $ \lim_{t\to+\infty}\langle A_i(t) B_j(t)\rangle = D^{(2)}_{AB}(x_i,x_j)$ only (cf. Appendix~\ref{app:sqt}), where
\begin{multline}
D^{(2)}_{AB}(x_i,x_j) =  -\delta_{ij} + 2 \sum_{k}
\varphi^*_k(x_i) \varphi_k(x_j) N_0(k),
\end{multline}
with $\varphi_k(x_i) = \sqrt{2/(L+1)} \sin k x_i$.
This result implies that $D^{(2)}(x_i,x_j)$ is nonergodic.
Hence, $C^{(2)}_{xx}(x_i,x_j,t\to+\infty)$ and $S(q,t\to+\infty)$ are also nonergodic.

\section{Discussion, Summary, and Outlook}

 We have shown that, in exactly solvable models,  the 
 correlation functions  of both local and non-local operators, at asymptotically long times,  
 are functions of the quasi-particle occupations in the initial state,  for a broad class of  
 initial states.  This means that correlation functions in these systems 
 retain much more memory of the initial conditions than in systems relaxation to thermal
 equilibrium, which is described by a standard Gibbs ensemble. 
 This lack of relaxation is similar to the observations of McCoy~\cite{McCoyII} 
 and Mazur~\cite{Mazur} for the magnetization in the XY model. It implies equilibration~\cite{Reimann} but
 lack of ergodicity~\cite{Mazur}, as the existence of non trivial integrals of motion
 strongly constraints the system dynamics and prevents it from reaching thermal equilibrium and 
 exploring all possible states having the same average energy and particle number (grand
 canonical Gibbs ensemble).
 
 By using the reduced density matrices for this class
 of initial states, we can show that nonergodicity implies that the asymptotic correlation 
 functions can be effectively described by an ensemble that assigns a mode-dependent
 temperature to each eigenmode. This is precisely the physical content of 
 the generalized Gibbs ensemble (GGE)~\cite{Rigol_gge}.
 We have illustrated our method by analyzing the
 the quantum Ising  and the XX spin chain models, both of which exhibit fermionic quasi-particles. In 
 Sec.~\ref{sec:LM}, the Luttinger model, which exhibits bosonic quasi-particles,
 was analyzed by the same method.

  For initial states lacking the  lattice translational symmetry, we have
 shown the connection between nonergodicity and the fact that eigenmode correlations in the initial state
 become dominated by diagonal correlations  (\emph{i.e.} quasi-particle occupations) as the  thermodynamic limit is approached.
By direct numerical calculation and  physical reasoning, we have argued that this property should hold true for
quantum quenches in which a physically sensible potential (e.g. a trap) that scatters the quasi-particles is suddenly 
removed at $t =0$. However, at present we are unable to provide a mathematically rigorous  proof of this fact (cf. 
Eq.~\ref{eq:important}),  although in all cases that we have examined so far, it appears to hold true. 

 Furthermore, using method reported here, we have been able to \emph{analytically} shed light,
for a much broader class of exactly solvable models and initial states than considered so far~\cite{Cazalilla,Calabrese_Ising},
on the conditions under which the generalized Gibbs ensemble is expected to apply. Thus, our results
extend  the validity of the GGE conjecture to a much broader class of quantum quenches. 
Our method also explains the special role played by the quasi-particle occupation operators as the set
of integrals of motion required to construct the GGE. The nonergodic behavior of the correlation functions
found here is entirely explained by the dependence on the expectation value of such 
integrals of motion alone.

 In future studies~\cite{unpub,unpub2}, it will interesting to understand how these results relate
to the generalized eigenstate thermalization hypothesis discussed in Ref.~\cite{Cassidy}.
We will also apply our methods  to understand the conditions under which the asymptotic
state can become arbitrarily close to a thermal state~\cite{unpub}. The latter study  unveils further
interesting connections between the GGE and quantum Information 
theory~\cite{unpub}.

\acknowledgments

 The authors thank M. Rigol, M. Olshanii, R. Fazio, and L. Amico, for  enlightening discussions,
A. Polkovnikov and G. Mussardo for a careful reading and useful comments on the manuscript,
and J. H. H. Perk for his remarks on the preprint and  for bringing  Refs.~\cite{Mazur,McCoyII}
to   their attention. MAC also thanks D.W. Wang for useful discussions and for his
kind hospitality at  NCTS (Taiwan)  and  acknowledges the support of
Spanish MEC grant FIS2010-19609-C02-02.
MCC acknowledges the NSC of Taiwan

\appendix

\section{Eigenmodes of the quantum Ising and XX chains}\label{app:eigmodes}

The Hamiltonian of the XX and Quantum Ising chains introduced in the
main text can be brought to diagonal form by means of
the Jordan-Wigner transformation:
\begin{align}
\sigma^z_i &= 1 - 2 f^{\dag}_i f_i,  \quad \sigma^{+}_i = \prod_{j < i} (1 - 2 f^{\dag}_j f_j) f_i,\label{eq:jw}\\
\sigma^x &=  \frac{1}{2}(\sigma^{+}_i + \sigma^{-}_{i}), \quad \sigma^{-}_i = \left(\sigma^{+}_i \right)^{\dag}
\end{align}
with $\{f_i, f^{\dag}_j\} = \delta_{ij}$, anti-commuting otherwise.

For the Quantum Ising chain, assuming periodic boundary conditions, that is,
\begin{equation}
f_j = \left(\frac{1}{L}\right)^{1/2} \sum_{k} \, e^{i k x_j} f(k),
\end{equation}
with $x_j = j$  and
\begin{equation}
f(k) = \cos (\theta_g(k)/2) \gamma(k) + i \sin (\theta_g(k)/2) \gamma^{\dag}(-k),
\end{equation}
where $\tan \theta_g(k) = \sin k/(\cos k - g)$.
These two transformations render the Hamiltonian of the Quantum Ising chain, Eq.~\ref{eq:hamqi}, 
diagonal:
\begin{equation}
H = \sum_k \epsilon_g(k) \left[ \gamma^{\dag}(k) \gamma(k)  - \frac{1}{2} \right],
\end{equation}
where $\epsilon_g(k) = 2 J\sqrt{1+g^2-2g^2 \cos k}$.

 For the XX chain we shall assume an open ended chain where
\begin{equation}
f_j = \left(\frac{2}{L+1}\right)^{1/2} \sum_{k} \sin k x_j\:  f(k),
\end{equation}
with   $k = \frac{\pi m}{L+1}$, $j,m=1, \ldots, L$, yields:
\begin{equation}
H = \sum_{k} \epsilon(k) f^{\dag}(k) f(k),
\end{equation}
where $\epsilon(k) = -J \cos k - h$.

\section{Time averages and Wick's theorem} \label{app:wick}

As mentioned in the main text, the calculation of asymptotic correlation functions
of non-local operators like $\sigma^x_j$ in the quantum Ising and XX chain models
depends on  the applicability of Wick's theorem in the $t\to +\infty$ limit to multi-point correlation functions.
Thus, we shall first tackle this problem by  time-averaging the correlation functions of finite systems prior
to taking the thermodynamic limit.  Let
\begin{equation}
C^{(2)}_O(x_i,x_j,t) = \langle O(x,t) O(0,t) \rangle,
\end{equation}
for local operator like $O(x) = \sum_k \varphi_k(x) f_k$.
Its time average is defined as:
\begin{equation}
\overline{C^{(2)}_{O}(x_i,x_j,t)} = \lim_{T\to +\infty} \frac{1}{T} \int^{T}_0 dt\,
C^{(2)}_O(x_i,x_j,t). \label{eq:timeav}
\end{equation}
\emph{A priori}, the time  average of $C^{(2)}_{O}(x_i,x_j,t)$  followed
by the thermodynamic limit yields
\begin{align}
D^{(2)}_{O}(x_i,x_j)  &= \sum_{k}\varphi^*_k(x_i)\varphi_{k}(x_j) N_0(k), \notag \\
 &= \lim_{t\to +\infty} C^{(2)}_O(x_i,x_j,t),
 \label{eq:coinf2}
\end{align}
where the last limit is taken after  the thermodynamic limit. 
However, taking the thermodynamic limit after time averaging
is a subtle procedure. For instance,  for the four-point correlation,
\begin{multline}
C^{(4)}_O(x_i,x_j,x_m,x_n,t) = \langle O^{\dag}(x_i,t) O^{\dag}(x_j,t) \notag \\
\times  O(x_m,t) O(x_n,t) \rangle,
\end{multline}
it yields:
\begin{multline}
\overline{C^{(4)}_O(x_i,x_j,x_m,x_n,t)}  = \sum_{k, k^{\prime}} A_{k,k^{\prime}}(x_i,x_j,x_m,x_n)\\
\times \left[ N_0(k) N_0(k^{\prime}) - |G_0(k,k^{\prime})|^2 \right],\label{eq:c4}
\end{multline}
where
\begin{multline}
A_{k,k^{\prime}}(x_i,x_j,x_m,x_n) = \varphi^*_{k}(x_i)
\varphi_{k}(x_n) \varphi^*_{k^{\prime}}(x_j) \varphi_{k^{\prime}}(x_m) \\
-  \varphi^*_{k}(x_i)
\varphi_{k}(x_m) \varphi^*_{k^{\prime}}(x_j) \varphi_{k^{\prime}}(x_n).
\end{multline}
Since we have assumed that the square of the functions $\varphi_k(x_i)$ is normalized to system size,
 \emph{i.e.} $|\varphi_k(x_i)| \sim O(L^{-1/2})$, it follows that $A_{k,k^{\prime}}(x_i,x_j,x_m,x_n)
 \sim O(L^{-2})$, which is required to obtain a finite result in the $L\to+\infty$ limit
 given the presence of the double sum over $k$ and $k^{\prime}$.

However, we note that
\begin{multline}
\overline{C^{(4)}_O(x_i,x_j,x_m,x_n,t)} \neq \overline{C^{(2)}_O(x_i,x_n,t)} \,\, \overline{C^{(2)}_O(x_j,x_m,t)} \\
- \overline{C^{(2)}_O(x_i,x_m,t)} \,\, \overline{C^{(2)}_O(x_j,x_n,t)}.\label{eq:nonwick}
\end{multline}
and thus, if we  also identify $D^{(4)}(x_i,x_j,x_m,x_n) = \lim_{t\to +\infty} C^{(4)}_O(x_i,x_j,x_m,x_n,t) $ with
its time average,  Wick's theorem appears to be violated for $t \to +\infty$ in the thermodynamic limit as
$D^{(4)}(x_i,x_j,x_m,x_n)$ will \emph{a priori} depend on all quantum correlations between the
eigenmodes described by $G_0(k,k^{\prime})$ (cf. Eq.~\ref{eq:c4}).  However, $D^{(2)}(x_i,x_j)$,
which we identified with $\overline{C^{(2)}(x_i,x_j,t})$,  only depends on $N_0(k) = G_0(k,k)$. This has implications for
the calculation of correlation functions of non-local operators in the $t\to +\infty$ limit.
Nevertheless, as it was discussed in the main text,  the  eigenmode
 correlations,
 \begin{equation}
 |G_0(k,k^{\prime})|^2 \to \left[ N_0(k)\right]^2 \delta_{k,k^{\prime}} + \Delta R_0(k,k^{\prime}),
 \end{equation}
 as the thermodynamic limit is approached. Therefore, the term involving these correlations
 in Eq.~\eqref{eq:c4}, becomes approximately equal (after neglecting $\Delta R_0(k,k^{\prime})$  to
 \begin{equation}
  -\sum_{k} A_{k,k}(x_i,x_j,x_m,x_n) N^2_0(k),
 \end{equation}
which is manifestly of $O(L^{-1})$ as $L\to +\infty$.
On the other hand, the term
\begin{equation}
 \sum_{k, k^{\prime}} A_{k,k^{\prime}}(x_i,x_j,x_m,x_n) N_0(k) N_0(k^{\prime}) \label{eq:dom}
\end{equation}
is of $O(L^0)$ as $L\to +\infty$. However, Eq.~(\ref{eq:dom}) equals the anti-symmetrized
product of the time averaged two-point correlation functions, $\overline{C^{(2)}(x_i,x_j)}$.
Thus, the fact that the contribution of non-diagonal correlations becomes negligible in the thermodynamic
limit justifies the procedure to taking the time-average followed by the thermodynamic limit.

 We can try to extend the  above result to higher order (\emph{i.e.} $3$-point, etc.)
correlation correlation functions. However, it is more convenient to take a shortcut.
As discussed in the  previous paragraph,
we can identify the time average of \emph{two-point} correlation functions of local
operators  like $O(x)$ with the $t \to +\infty$ limit
of the same correlation function in the thermodynamic limit.  These two point
correlation function are the building blocks for computing with multi-point correlation functions
or correlation functions of non-local operators like $\sigma^x$ as we can
apply Wick's theorem first and then let $t \to +\infty$ using
\begin{equation}
\lim_{t \to +\infty} c_1(t) c_2(t) \cdots c_M(t) = \prod_{i=1}^M c_{i}(+\infty)
\end{equation}
where $c_i(+\infty) = \lim_{t\to +\infty} c_i(t)$ since the  $t \to +\infty$ limit of
two-point every correlation function ($c_i(t)$ in the expression above)
exist and it is given by the its time average followed by the thermodynamic limit.
\section{Non-local correlations in the quantum Ising and XX chain}\label{app:sqt}
 For the XX chain, the momentum distribution of the hardcore bosons can be obtained from
 the expression:
\begin{align}
S(q,t)
=   \frac{1}{2L}
\sum_{ij}  e^{i q (x_i-x_j)}\:  C^{(2)}_{xx}(x_i,x_j,t), \label{eq:sq}
\end{align}
for $t\to +\infty$ can be computed first using Wick's theorem and letting $t\to+\infty$ in the resulting expression.
To this end, it is convenient to write  $\sigma^x_i(t) = A_i(t) \prod_{j<i} A_j(t) B_j(t)$, being
\begin{align}
A_i(t) &= f^{\dag}_i(t) + f_i(t)  = \left(\frac{2}{L+1}\right)^{1/2} \sum_{k} \sin k x_i \: \big[
e^{i\epsilon(k) t/\hbar} \notag \\
&\times  f^{\dag}(k) + e^{-i\epsilon(k) t/\hbar} \: f(k)\big], \\
B_i(t) &= f^{\dag}_i(t) - f_i(t) =  \left(\frac{2}{L+1}\right)^{1/2} \sum_{k} \sin k x_i \: \big[
e^{i\epsilon(k) t/\hbar}  \notag\\
 &\times f^{\dag}(k) - e^{-i\epsilon(k) t/\hbar} \: f(k)\big].
\end{align}
where the mode expansions are given for the XX chain (cf. e.g. Ref.~\cite{Sengupta04} for the corresponding
expressions for the quantum Ising chain).  Hence,
\begin{equation}
C^{(2)}_{xx}(x_i,x_j,t) = \langle B_i(t)\left[ \prod_{i<l <j} A_l(t) B_l(t) \right] A_j(t) \rangle,
\end{equation}
where we have assumed that $x_i < x_j$ without loss of generality.
To evaluate the expression above we use  Wick's theorem and take the limit $t \to +\infty$
of the resulting expression only after taking the thermodynamic limit.  Moreover,
since
\begin{align}
\langle A_i(t) A_j(t) \rangle &\to \delta_{ij},\\
\langle B_i(t) B_j(t) \rangle &\to -\delta_{ij},
\end{align}
the correlation function
\begin{equation}
D_{xx}(x_i,x_j) = \lim_{t\to +\infty} C_{xx}(x_i,x_j,t)
\end{equation}
can be written  as a Toeplitz  determinant~\cite{McCoyI}:
\begin{align}
D_{xx}(x_i,x_j) &=&  \left|
\begin{array}{cccc}
a_0 & a_1 & \cdots & a_{-n+1}\\
a_1 &  a_0 & \cdots & a_{-n+2}\\
\vdots & \vdots & \ddots & \vdots \\
a_{n-1} & a_{n-2} & \cdots & a_0
\end{array}
\right|,\label{eq:toeplitz}
\end{align}
where $a_{i-j+1} = -D^{(2)}_{AB}(x_i,x_j)$, and
\begin{multline}
D^{(2)}_{AB}(x_i,x_j) =  \lim_{t\to+\infty} \langle A_i(t) B_j(t) \rangle \\
=  -\delta_{ij} + 2 \sum_{k}
\varphi^*_k(x_i) \varphi_k(x_j) N_0(k), \label{eq:dd}
\end{multline}
where  $\varphi_k(x_i) = \sqrt{2/(L+1)} \sin k x_i$ and
the thermodynamic limit $L\to +\infty$ at finite lattice filling, $n_0 = N /L$,
is implicitly understood. Note that, in this limit, the actual boundary conditions (open
or otherwise) are irrelevant.

\end{document}